\pdfoutput=1
\documentclass[sigconf]{acmart}

\usepackage{adjustbox}
\usepackage{colortbl}

\usepackage[framemethod=TikZ]{mdframed}
\usepackage{xcolor} 
\definecolor{quote}{HTML}{9673A6}

\newcommand{\pquote}[1]{\textcolor[HTML]{9673A6}{\textbf{Participant #1:} }}

\newenvironment{zeroindent}
  {\par\setlength{\parindent}{0pt}}
  {\par}

\definecolor{customcolor}{HTML}{6C8EBF}

\newenvironment{fancy}[2][]{
    \mdfsetup{
        skipabove=1pt, 
        innerlinewidth=1pt, innerlinecolor=#2, 
        linewidth=0pt,
        backgroundcolor=#2!20 
    }
    \begin{mdframed}}
    {\end{mdframed}}

\newenvironment{coloredframe}[2][]{
    \mdfsetup{
        skipabove=2pt, 
        hidealllines=true, leftline=true,      
        innerlinewidth=2pt, innerlinecolor=#2, 
        linewidth=0pt,
        backgroundcolor=#2!10
    }
    \begin{mdframed}}
    {\end{mdframed}}

\newcommand{\dialoguegpt}[2]{
    \begin{coloredframe}{#1}
    \vspace{1px}
    \small 
    \begin{zeroindent} #2 \end{zeroindent}
    \vspace{1px}
    \end{coloredframe}
    \vspace{-5px}
}

\acmConference[FSE '25]{Companion Proceedings of the 33rd ACM Symposium on the Foundations of Software Engineering}{June 23--27, 2025}{Trondheim, Norway}

\acmBooktitle{Companion Proceedings of the 33rd ACM Symposium on the Foundations of Software Engineering (FSE '25), June 23--27, 2025, Trondheim, Norway}

\begin{document}

\title[Factors Influencing Gender Representation in IT Programmes]{Factors Influencing Gender Representation in IT Faculty Programmes: Insights with a Focus on Software Engineering in a Nordic Context}

\author{Cristina Martinez Montes}
\email{montesc@chalmers.se}
\orcid{0000-0003-1150-6931}
\affiliation{%
  \institution{Chalmers University of Technology and University of Gothenburg}
  \city{Gothenburg}
  \country{Sweden}
}
\author{Jonna Johansson}
\email{gusjohjoid@student.gu.se}
\orcid{0009-0003-1550-7883}
\affiliation{%
  \institution{Chalmers University of Technology and University of Gothenburg}
  \city{Gothenburg}
  \country{Sweden}}

\author{Emrik Dunvald}
\email{gusdunvem@student.gu.se}
\orcid{0009-0002-5511-1800}
\affiliation{%
   \institution{Chalmers University of Technology and University of Gothenburg}
  \city{Gothenburg}
  \country{Sweden}}


\begin{abstract}
Software engineering remains male-dominated despite efforts to attract and retain women. Many leave the field due to limited opportunities, unfair treatment, and challenging workplace cultures. Examining university life and choices is important, as these formative experiences shape career aspirations and can help address the root causes of underrepresentation in the industry.
The study aimed to deepen understanding of the motivations behind women's choice of a career in IT, their experiences in academic life, and how these experiences influence their career decisions, all within a Nordic context. We used a combination of surveys in the bachelor programmes in the IT faculty and interviews with only women from software engineering (SE) to provide a comprehensive view of population experiences and a closer exploration of the experiences of a smaller sample with a focus on SE. 

Our results showed that family and personal interest are among the main factors motivating women to choose an IT programme. Further, women perceive more challenges following their chosen career path than men. We proposed high-level actions to address gender-related challenges and disparities based on our findings.
\end{abstract}




\maketitle

\section{Introduction} \label{sec:intro}

The representation of genders within the IT industry and educational settings has been extensively researched, revealing significant imbalances and challenges. Nordic countries are often regarded as leaders in gender equality \cite{larsen2021gender}; however, they continue to face challenges in achieving gender balance in IT education and careers. 

In Sweden, for instance, the field of software development demonstrates a notable gender gap, with a higher percentage of male employees than women \cite{scb2021}. This gender disparity extends beyond Sweden, as noted by Zhu \cite{zhu2019recoding} and Atal et al. \cite{atal2019gender}, indicating a global trend of male dominance within the IT industry.

Despite significant efforts to encourage women to enter the industry in Sweden, the challenge lies in retaining them \cite{tokbaeva2023career}. Discrimination within the Swedish IT industry is often discreet and hidden, requiring women to demonstrate resilience \cite{zhu2019recoding, tokbaeva2023career, montes2024qualifying} continually. Examples include women being assigned traditionally feminine roles such as front-end developers or designers \cite{zhu2019recoding}. Additionally, many women face early career pressures to shift towards managerial positions rather than pursuing technical roles \cite{davies2005gender}.
Women studying or working in IT often encounter challenges unique to their gender that others do not \cite{zhu2019recoding, tokbaeva2023career, davies2005gender}. 

Other nordic countries face similar situations. Corneliussen~\cite{corneliussensuperpowers} interviewed Norwegian women and identified factors that motivated them to pursue IT careers, including their strengths in related fields such as science, logic, and languages, which gave them confidence and a sense of familiarity. Many viewed programming as akin to learning a new language or were drawn by broader interests, such as curiosity about technology's role in the world. Lacking clear plans, they often took unplanned routes into IT, with opportunities and exploration guiding their decisions. Other factors included the influence of female role models, and for women from Southern Europe and Asia, the motivation came from viewing IT as a suitable field for women \cite{corneliussen2024reconstructions}.

Although gender representation in IT careers has been widely studied, the role which education has in this issue is less researched, particularly in the context of Sweden. Our study aims to explore gender inequality within an educational context, focusing on identifying factors influencing gender representation in IT-related programmes at the University of Gothenburg. It also explores how women's experiences in these programmes shape their academic journeys, career aspirations, and perceptions of gender-related challenges in the IT field.

By doing so, we seek to gain insights into factors contributing to gender disparities and identify potential strategies for creating a more inclusive educational environment. These insights can help inform efforts to improve gender balance and reduce barriers to women entering and thriving in the IT field.

This study looks at the programmes in the IT faculty of the University of Gothenburg that represent diverse focus areas, from those centred on management, design, and user experience to those with a stronger emphasis on technical aspects, such as hardware and back-end development.

The research questions below guided this study by focusing on the representation of lived experiences and career-related perceptions shaped within these educational contexts.

\textbf{RQ1: What factors influence gender representation among students in IT-related programmes?}

This RQ aims to identify barriers and opportunities shaping women's entry into IT, informing efforts to improve gender diversity in education.

\textbf{RQ2: How are women's experiences in IT-related programmes regarding gender?}

This RQ explores women's gender-specific challenges in IT education, helping identify areas for creating more inclusive academic environments.

\textbf{RQ3: How does the university experience shape career aspirations and perceptions of gender-related challenges in IT fields?}

We aim to examine how university experiences influence women's career aspirations and perceptions of gender-based obstacles in IT, guiding support strategies.

We conducted surveys to get general knowledge about the reason for choosing IT programmes and the experiences at the university. We conducted interviews to go deeper into the general knowledge from the survey.
\section{Background} 
\label{sec:background}
This section outlines the research context and points out the knowledge gap that our study seeks to address.

\subsection{Factors and Strategies Influencing Women's Pursuit of STEM Education}

Research on women's choices in higher education has shown several factors that shape their decisions, particularly in the context of STEM fields. Støren and Arnesen \cite{storen2007women} identified parental influence and social status as significant determinants, with women from higher social classes more inclined to pursue male-dominated fields. Additionally, women who excel in mathematics tend to choose more gender-neutral disciplines. At the same time, social factors—such as the desire for meaningful work, compatibility with family responsibilities, and concerns about being a minority in the workplace—tend to guide their decisions more than individual strengths or weaknesses, which are a greater focus for men \cite{malik2018social}. Furthermore, Cheryan et al.\cite{cheryan2017some} highlight three main factors: limited early exposure to STEM, gender disparities in self-efficacy, and the male-dominated culture in certain STEM fields. This includes stereotypes about who typically participates in these fields, assumptions regarding women's ability to succeed, and the scarcity of positive female role models. 

Additionally, gendered stereotypes about STEM fields significantly influence career choices. For instance, women who encounter computer science majors that fit the "nerdy" stereotype are less likely to pursue the field \cite{cheryan2013enduring}. Women also tend to prefer people-oriented careers, which helps explain their higher participation in social and life sciences compared to engineering or physics \cite{rong, yang2015gender}. A UK study also found that female students' lack of interest in technical details and desire to make a positive impact was negatively correlated with their intention to pursue physics, possibly explaining their engagement in life sciences \cite{mujtaba2013sort}.

To foster greater female participation in STEM, Wang and Degol~\cite{wang2017gender} emphasise the importance of nurturing interest in these subjects early in education. Encouraging young women to develop an interest in STEM from middle school onwards is important, particularly for those with talent but little initial interest in these fields. The authors suggest that promoting achievement in STEM subjects at school can cultivate long-term interest and help close the gender gap in these areas.
Further, Tandrayen-Ragoobur and Gokulsing~\cite{tandrayen2022gender} offer several strategies to encourage young women to engage with STEM. Teachers have a crucial role in challenging stereotypes and motivating students by incorporating discussions about female role models in STEM. Parental support also proves essential, as women who receive encouragement from their families are more likely to pursue STEM education. Moreover, activities such as STEM clubs, internships, and mentorship programs can effectively foster greater interest and involvement in these fields.

\subsection{Getting Women to Stay in the Industry}

While considerable efforts are directed towards encouraging women to enter the IT industry, a more significant challenge lies in ensuring their retention. Tokbaeva and Achtenhagen \cite{tokbaeva2023career} found that women in the industry often experience a sense of isolation, lacking meaningful support, which hinders their long-term engagement. Despite initiatives to improve retention, these women report little progress in tangible benefits or improved experiences. The authors suggest focusing on how women cope with the challenges of gendered workplaces, particularly the resilience they must continually practice in response to everyday subtle discrimination.
Zhu \cite{zhu2019recoding} highlighted the experiences of immigrant women in Sweden's programming sector, who are drawn to the industry by the opportunity to enter the labour market without language barriers and see programming as a means to achieve career goals that align with their interests. However, these women face challenges such as being stereotyped into traditionally feminine roles, like front-end development or design. They experience discrimination in more covert forms and must cultivate career resilience to overcome these hurdles. Another issue is that women, especially in the early stages of their careers, are often encouraged to pursue management roles over technical ones. However, they typically remain in middle-management positions, while men still dominate top leadership roles. This could be influenced by assumptions about women's interpersonal skills, a reluctance from men to compete for technical positions, or even quota-driven appointments aimed at meeting gender equality targets \cite{davies2005gender}.

\subsection{Sweden as Context}

Sweden has addressed barriers to women's career advancement for over 50 years \cite{holst2001institutionelle, winn2004entrepreneurship}, yet women remain underrepresented in IT. While 19\% of software and system developers are women, this profession remains one of the top 10 most common for men~\cite{scb2021}. In 2018, only 29\% of new engineering students were women, which had not changed over the previous five years \cite{scb2022}. This contrasts sharply with the 86\% of girls who express interest in technology-oriented subjects at a younger age \cite{SchoolInspectorate2014}, suggesting a disconnect between early interest and career progression. Further evidence from Sweden reveals that women in IT often remain in entry-level roles \cite{tokbaeva2023career}, with various social and structural factors influencing their career choices, persistence, and advancement at different stages \cite{armstrong2014barriers, armstrong2018advancement, montes2025factors}. This demonstrates the challenges women face in transitioning from education to higher-level positions in IT despite early enthusiasm.

This study seeks to understand and fill a literature gap on the persistent gender disparities in IT education and careers. This study's results can help develop strategies to promote gender equality, enhance women's experiences in IT education, and ultimately increase their representation and success in IT careers.
\section{Methodology} \label{sec:methodology}

The data collection was done with surveys and semi-structured interviews. We used surveys to find overarching trends and identify patterns and correlations within a specific population \cite{linaaker2015guidelines} to help us understand the broader context of the issue. We complemented the survey data with interviews to get more breadth and depth data~\cite{segal2006structured} and explore the nuances of individual experiences, subjective perspectives, and contextual factors. By using this mixed-method approach, the study got scale and sensitivity, which aligned with our aim to investigate general patterns and individual-level variation.

\subsection{Population}
The study population consisted of bachelor’s level university students, with a primary focus on women, enrolled in all bachelor programmes of the IT faculty at the University of Gothenburg: Datavetenskap (Computer Sciences), Systemvetenskap (Systems Science), Kognitionsvetenskap (Cognition Science), and Software Engineering and Management. We used purpose sampling since we targeted specific programmes in the university. We chose these programs because they are relevant to our research questions and provide insights from a diverse range of IT students since they vary in technical focus. The Datavetenskap programme is the most technically oriented, strongly emphasising programming and mathematics. In contrast, Kognitionsvetenskap focuses more on the interaction between humans and machines. The other two programmes offer a broader and more diverse range of courses.

\subsection{Data Collection}
This study used a cross-sectional survey that included individuals of all genders, enabling comparisons across genders for specific questions. Furthermore, the interviews were done only with women aiming to capture only their experience. Next, we elaborate on both instruments.

\subsubsection{\textbf{Survey}}

The survey was based on insights from Wolff et al. \cite{wolff2020prevents}, which discusses women in software engineering roles and other relevant literature. The questions were divided into four parts: 

\begin{enumerate}
    \item Demographic questions. 
    \item Factors influencing participants' choice of university programmes: We investigated whether gender distribution in university programmes \cite{lewis2016don} and the level of math and programming required \cite{wang2017gender} influenced participants' choices. Another key factor is parents' occupation and education, which studies like Støren et al. \cite{storen2007women} and Brooks \cite{brooks2003young} show can impact young men and women differently, depending on whether it is the mother or father. 
    \item Experiences of discrimination in IT education: These questions are informed by insights from the research conducted by Tokbaeva and Achtenhagen \cite{tokbaeva2023career} and Zhu \cite{zhu2019recoding}. Both studies analised various facets of discrimination experienced by women in the IT industry.
    \item Career Aspirations: These questions are based on Zhu \cite{zhu2019recoding}, which suggests women often feel pressured to shift toward frontend development and design, and Davies and Mathieu~\cite{davies2005gender}, which discusses how women are encouraged to move from technical roles to management positions.
\end{enumerate}

The survey was done using Google Forms and distributed online. We piloted the survey with a small sample to ensure that the questions were clear and compelling. 

We distributed the survey through private channels, including Discord, email, and direct messages on Instagram, as well as through personal invitations to students in the selected courses. In addition, teachers of the selected courses were emailed and asked to share the survey with their students. See the complete survey in our replication package.

\subsubsection{\textbf{Interviews}}

Semi-structured interviews with women from the software engineering programme were used to collect qualitative data, allowing exploration of insights beyond survey questions~\cite{segal2006structured, aksu2009questionnaries}. The interviews were recorded and transcribed to retain the data as accurately as possible. 

The interview guide was created after the survey as it was intended to support the survey. It is partially based on a similar interview from Hand et al. \cite{hand2017exploring}, where they structured the questions based on the general themes of their survey questions. The survey questions were grouped into six themes: motivation for programme choice, technicality associated with the programme, gender's effect on the study experience, gender representation in the programme, career issues, and family experience. These themes were designed to guide interviews in exploring ideas beyond the survey.

The interview guide was pilot-tested and revised to ensure quality before finalisation. For the complete list of questions, refer to our repository~\cite{zenodo2024data}. We contacted women students from the selected programmes using various methods, including having a link in the survey, personal invitations and private messaging on Discord. The interviews were recorded with the consent of the participants and later transcribed to be analysed.

\subsection{Data Analysis}
We analysed the \textbf{survey} data using RStudio to create the visualisation plots to clearly depict the distribution of responses across different categories, highlight trends and patterns, and facilitate comparisons between subgroups. Subsequently, we analysed the \textbf{interviews} using thematic analysis by Braun and Clarke's six-phase approach \cite{braun2021thematic}. The analysis was done manually between two researchers. Initially, the data were transcribed verbatim to ensure accuracy and completeness. Then, we read the transcripts to familiarise ourselves with the data. Subsequently, two researchers independently coded one interview (20\% of the data) and measured inter-coder reliability (ICR) using Cohen's kappa (k), as recommended by Lombard et al. \cite{lombard2010intercoder}. The ICR was calculated to be 0.804, indicating an 80\% agreement across all coded data, which was deemed acceptable. Later, the transcripts were unitised, and both researchers coded all interviews separately. The next phase involved reviewing the codes and transcripts collaboratively to identify and define themes. After the first themes were finalised, the three researchers reviewed and refined them, leading to the final step, writing the findings.  

\subsection{Ethical Considerations}
This research followed the informed and voluntary consent guidelines outlined in GDPR Article 7 \cite{GDPR2018}. We informed participants how we would use the data, explained that they could withdraw from the study anytime and asked them to sign informed consent forms. Additionally, we anonymised the data when analysing it.

\subsection{Data Availability} 

We make the survey, interview guide, and survey answers available in our repository \cite{zenodo2024data}. The interview transcripts contain private information about our participants. Hence, we only shared the quotes in the paper to protect the participants' confidentiality and privacy.

\section{Results} \label{sec:results}
This section presents the results of the survey and the interviews and answers our research questions.

\subsection{Quantitative Findings}

We received 119 survey responses from students in the first, second, and third years of their bachelor’s programme. Table~\ref{tab:respondents} presents the distribution of the respondents by gender and programme.

\begin{table}
  \caption{Survey Respondents}
  
  \label{tab:respondents}
  \small
  \begin{tabular}{lccc}
    \toprule
    Programme & Women & Non-binary & Men \\
    \midrule
   Software Engineering and Management & 16 & 1 & 25 \\
   Cognition Science & 30 & 0 & 6 \\
    Systems Science & 15 & 0 & 11\\
    Computer Sciences & 7 & 2 & 6 \\
    \bottomrule
  \end{tabular}
\end{table}

\subsubsection{\textbf{RQ1 Factors Influencing Gender Representation}}

\begin{figure}
    \centering
    \includegraphics[width=1.05\linewidth]{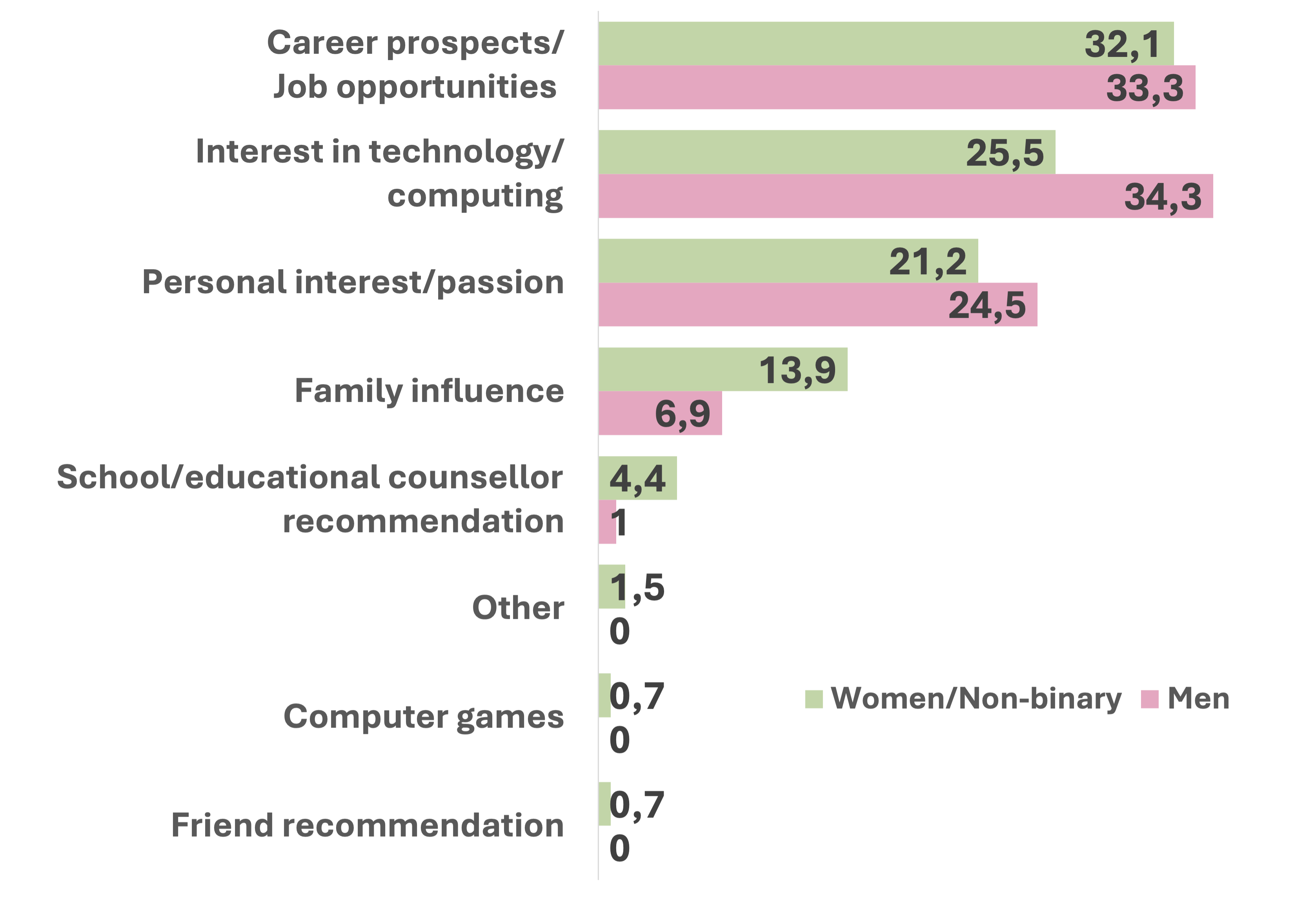}
    \caption{Answers to the question "What influenced your decision to pursue a degree in an IT-related field?" divided by gender. Participants could choose more than one factor.}
    \label{fig:Q4}
\end{figure}

We asked students what factors influenced their programme choice. Figure~\ref{fig:Q4} shows the answers grouped by women/non-binary individuals and men. The percentages reflect how often each factor was mentioned as influential in their decision to pursue a degree in an IT-related field.

\textbf{Career prospects/job opportunities} were the most frequently mentioned factor, accounting for 32.1\% of the mentions by women/\hspace{0pt}non-binary individuals
 and 33.3\% by men. 
\textbf{Interest in technology/ computing} was another leading factor, with 25.5\% of mentions from women/non-binary individuals and 34.3\% from men. However, men placed more emphasis on technology/computing, while women/non-binary individuals are more influenced by family and educational recommendations. Other personal influences, such as computer games or friend recommendations, were minimal in the decision-making process.

\begin{figure}
    \centering
    \includegraphics[width=1\linewidth]{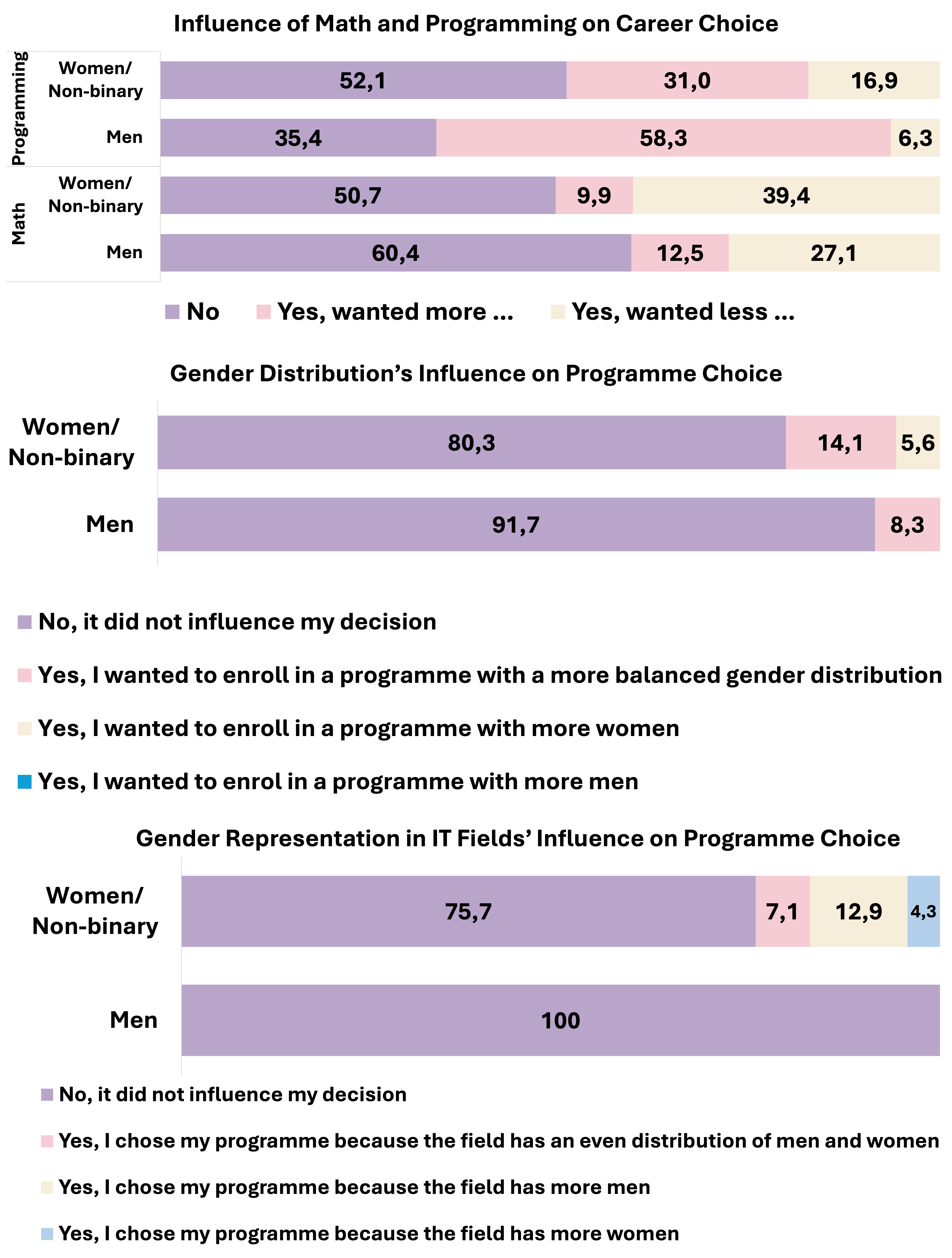}
    \caption{Factors Influencing IT Programme Selection}
    \label{fig:Q6-Q9}
\end{figure}

Regarding the influence of programming on program selection, the analysis revealed a clear gender divergence, see Figure~\ref{fig:Q6-Q9}. Most women and non-binary (52\%) reported that the amount of programming did not significantly affect their decision, while men (58\%) preferred programmes with a greater emphasis on programming. As for mathematics, both groups showed similar responses. Most men (64.4\%) indicated that the amount of mathematics did not influence their decision. Similarly, 50.7\% of women and non-binary respondents agreed with this view.

Concerning gender distribution, the responses from both groups were similar. A significant percentage (80.3\%) of women and non-binary respondents indicated that gender distribution did not influence their programme choice, while 91.7\% of men shared the same answer. Furthermore, when assessing the role of gender representation in program selection, a distinct pattern surfaced. All men respondents indicated that gender representation was not a decisive factor in their decision-making process. In contrast, women respondents displayed more varied perspectives, though most (80\%) agreed that gender representation did not significantly influence their programme choice.

This result shows that men prefer programs with more programming content, while women are less influenced by this factor. Meanwhile, mathematics was not a key factor in selecting programmes for women, non-binary, and men. Additionally, gender distribution and gender representation are generally not significant factors in programme selection for either gender, although women have more diverse opinions on its importance.


\begin{figure}
    \centering
    \includegraphics[width=1\linewidth]{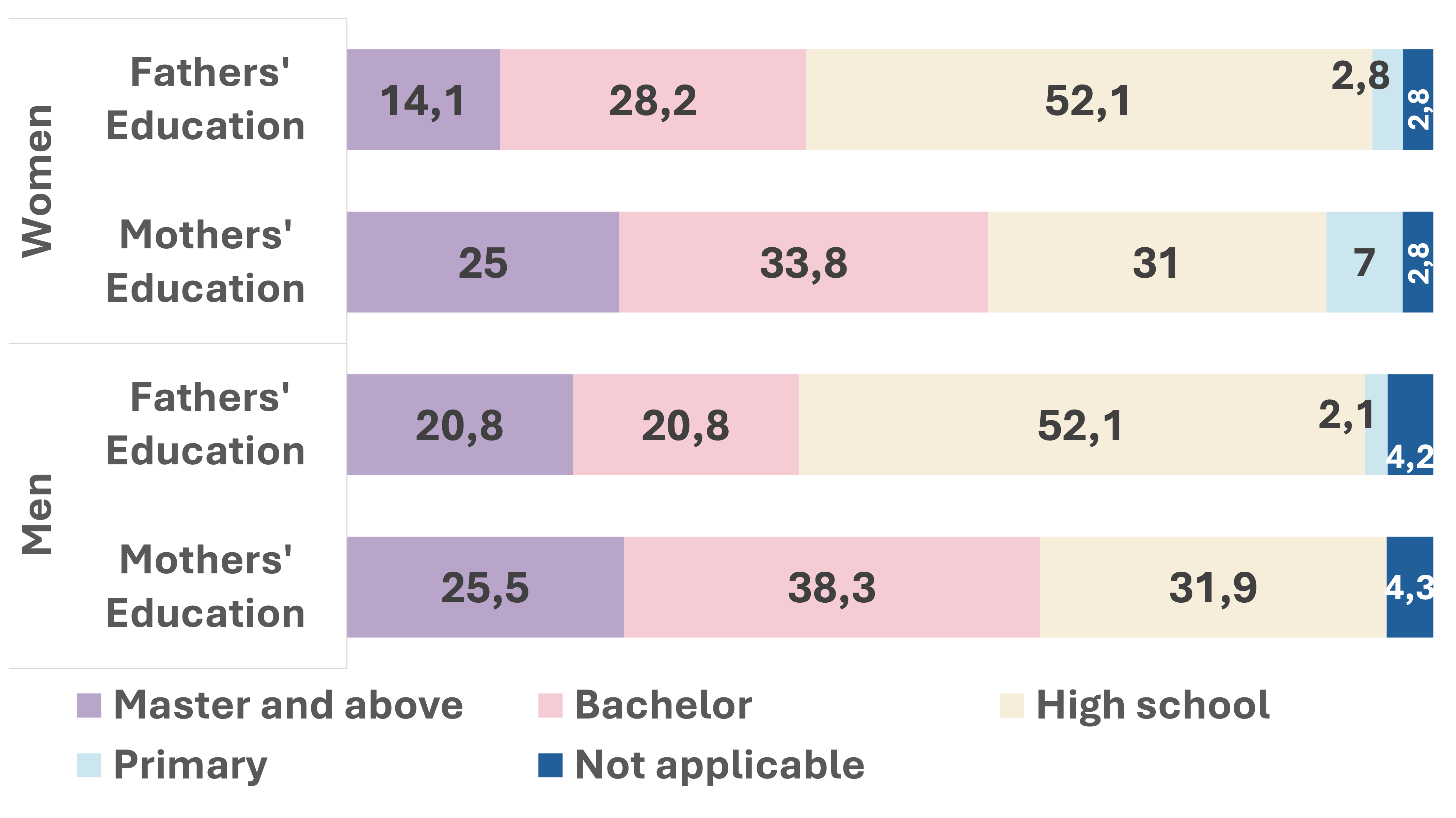}
    \caption{Parent's Education Level of Respondents}
    \label{fig:parents}
\end{figure}

When it comes to the influence of parents' education and occupation, Figure~\ref{fig:parents} illustrates the education levels, separated by respondents' gender

\textbf{Women and non-binary answers'} concerning fathers' occupations with high-level education accounted for 23.9\% of responses, including senior/leading occupations, legal and administrative roles, and positions like architect and teacher in IT. Technical education comprised 21.1\%, featuring IT-related technical jobs, nursing, and engineering. Service occupations requiring high school or some post-secondary training were reported by 19.7\%, covering roles like customer support and farming. Finally, 21.1\% marked "not applicable," reflecting absence, death, or lack of information, while 14.1\% listed unique roles such as builder, climbing instructor, or "retired."

About mothers' occupations, high-level education (university degree or higher) was reported by 40.8\%, including roles in legal, economics, teaching, dentistry, and senior occupations. Technical education accounted for 15.5\%, featuring technical IT roles, healthcare, and store management. Service occupations requiring high school or some post-secondary training were noted by 16.9\%, including service workers and cleaners. "Not applicable" was 22.5\%, while 2.8\% cited roles such as housewife and museum communication.

\textbf{Men answers'} regarding mothers' occupations, 25\% of the respondents reported roles requiring high-level education, such as air traffic controller, anesthesiologist, designer, head of HR, legal, economics, or administrative fields, and teaching. Skilled trades or technical education accounted for 16\%, including nurses and technical roles. Service occupations requiring high school or some post-secondary training were reported by 25\%, including preschool teachers, store salespeople, and other service jobs. Lastly, 30\% marked "not applicable," reflecting absence, death, or lack of information.

Concerning fathers' occupations for high-level education (university degree or higher), 37.2\% held roles like anesthesiologist, CTO, air traffic controller, legal, economics, and senior leadership positions. Skilled trades and technical education (vocational or college) comprised 11.6\%, including carpenters and technical IT-related roles. Service occupations, typically requiring high school or post-secondary education, accounted for 18.6\%, with service workers and sales roles. Lastly, 27.9\% marked "not applicable," reflecting absence, death, or lack of information.

\subsubsection{\textbf{RQ2 Women and Non-binary's Academic Experiences}}
\begin{figure}
    \centering
    \includegraphics[width=1\linewidth]{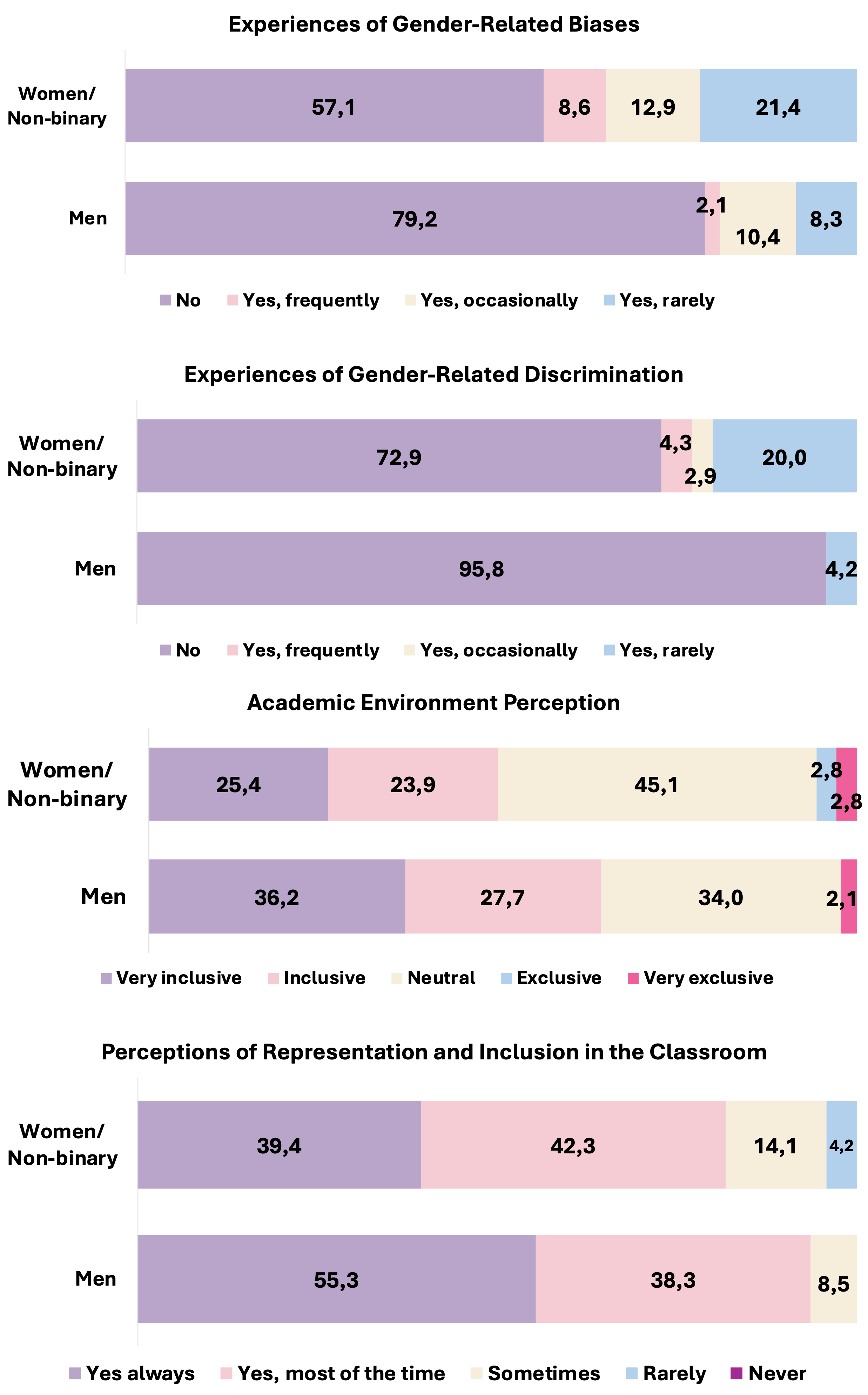}
    \caption{Experiences and Perceptions of Gender Inclusivity and Representation in IT Programmes}
    \label{fig:RQ2}
\end{figure}

We compare the experiences of women and non-binary people to men in areas such as biases, discrimination, academic environment perception and representation and inclusion. Figure~\ref{fig:RQ2} shows the visualisation of their answers.

 Concerning experiences of gender-related bias, a majority of men (79.2\%) reported no experiences and their instances of bias were rare, with 10.4\% experiencing it "occasionally" and 8.3\% "rarely". Meanwhile, only 57.1\% of women reported no bias, significantly lower than men. Additionally, 42.9\% experienced bias, with 8.6\% stating it occurred "frequently," 12.9\% "occasionally," and 21.4\% "rarely".
Regarding experiences of gender-related discrimination, the vast majority (95.8\%) of men reported no experiences of gender-related discrimination, with only 4.2\% experiencing it "rarely". In comparison, 72.9\%  of women reported no discrimination; however, 27.1\% indicated they had experienced it. This includes 4.3\% who reported it "frequently," 2.9\% "occasionally," and 20\% "rarely".

For perceptions of academic environment inclusivity, men (36.2\%) perceive the environment as "very inclusive," and 27.7\% rate it as "inclusive". However, 34\% view it as "neutral," with minimal responses for "exclusive" (0\%) or "very exclusive" (2.1\%).
Fewer women and non-binary (25.4\%) describe the environment as "very inclusive," and 23.9\% rate it as "inclusive". A larger proportion (45.1\%) perceive it as "neutral," and a small percentage feel the environment is "exclusive" (2.8\%) or "very exclusive" (2.8\%).

Finally, for representation and inclusion in classroom discussions and group projects, women and non-binary are less likely to feel "always" represented and included (39.4\%) compared to men. However, 42.3\% feel included "most of the time". A notable portion (14.1\%) experience inclusion "sometimes," with 4.2\% stating they "rarely" feel included. On the contrary, the majority of men (55.3\%) feel "always" represented and included, with an additional 38.3\% indicating they feel included "most of the time". Only 8.5\% responded "sometimes," and no men reported feeling excluded ("rarely" or "never").

In summary, women and non-binary are significantly more likely than men to encounter gender-related bias and discrimination within their programs. Men view the academic environment as more inclusive than women and non-binary, many of whom perceive it as neutral or even exclusive. Men generally feel more consistently included in academic settings than women and non-binary, who report more sporadic feelings of inclusion.

\subsubsection{\textbf{RQ3 University Experiences Shaping Career Aspirations}}

We asked students about their plans to change careers based on their university experience. 

\begin{figure}
    \centering
    \includegraphics[width=1.05\linewidth]{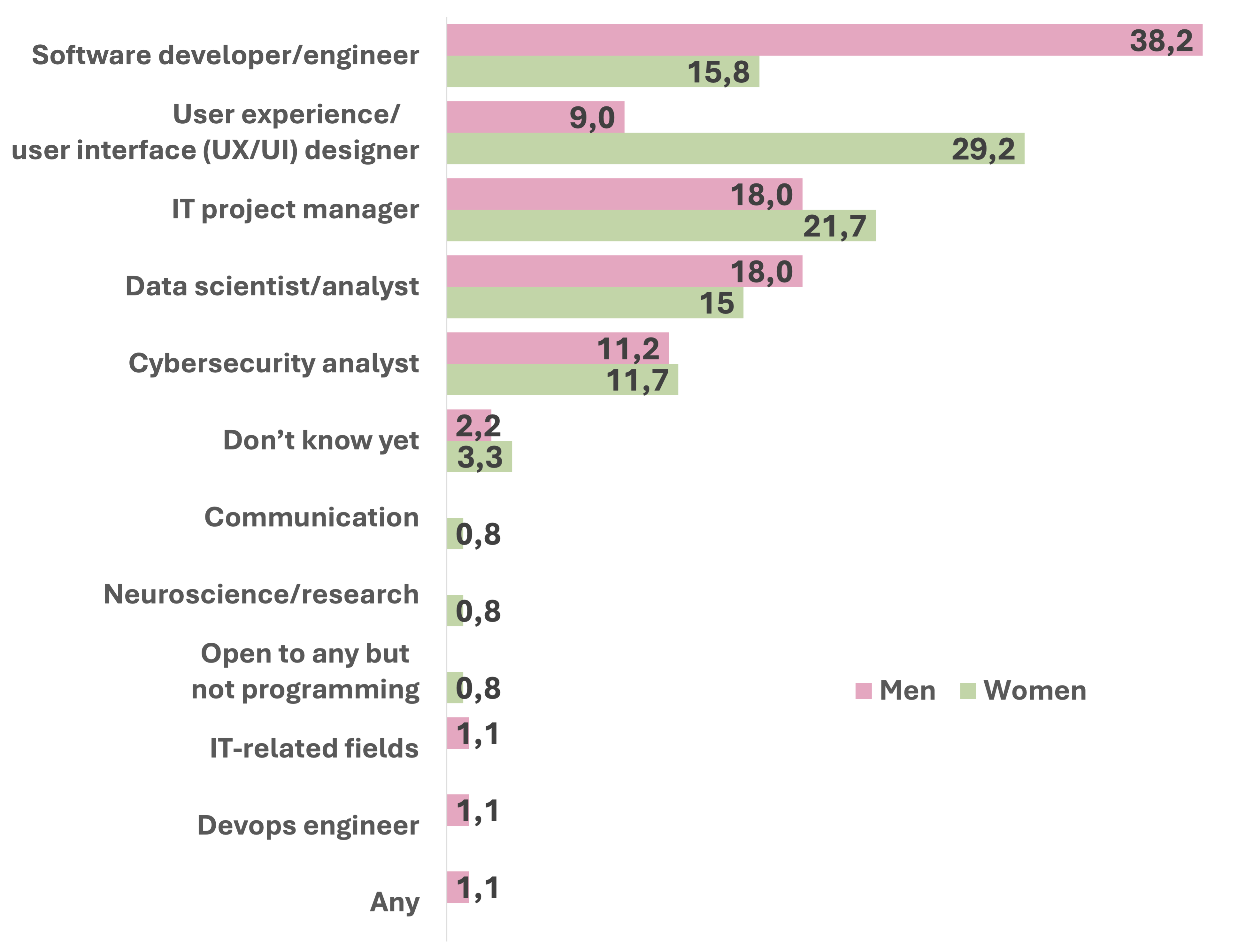}
    \caption{Career aspirations after completing IT degree}
    \label{fig:Q13}
\end{figure}

Figure~\ref{fig:Q13} shows the answers of students about their career plans after finishing university. A larger percentage of men plan to work as software developers/engineers (38.2\%) compared to women (15.8\%). Women are more likely to pursue user experience/user interface (UX/UI) design roles (29.2\%) than men (9\%). IT project management has a relatively balanced representation (18\% of men and 21.7\% of women). Other roles, such as data scientists/analysts and cybersecurity analysts, show similar participation between genders. A small portion of both men and women are undecided or open to roles outside of programming. It is important to note that the differences in curricula across the programmes (see table ~\ref{tab:respondents}) may influence the career aspirations reported in Figure 5. For instance, students in Software Engineering and Management may be more inclined toward software development, while those in Cognition Science may favour UX/UI design or research.



\begin{figure}
    \centering
    \includegraphics[width=1\linewidth]{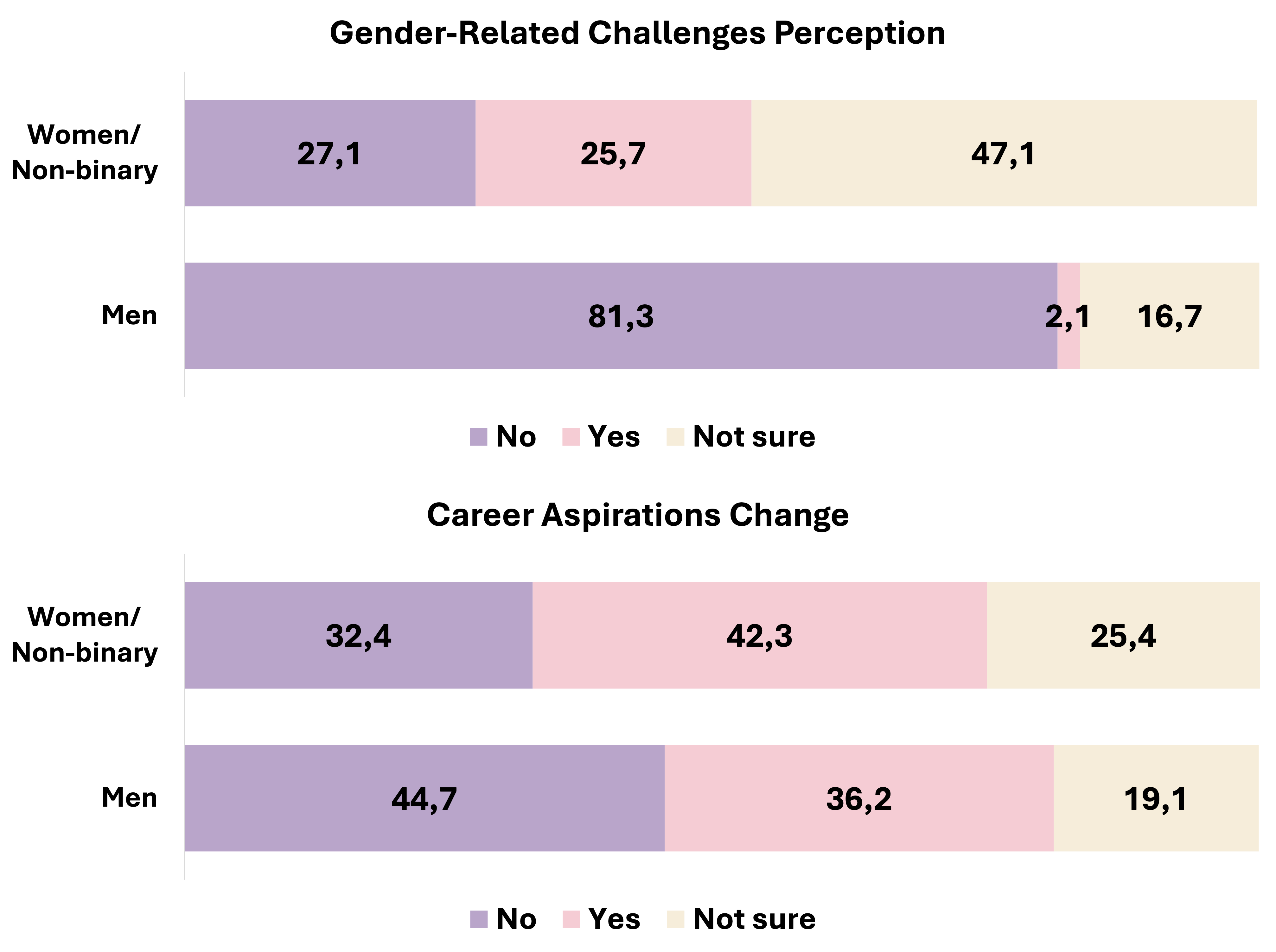}
    \caption{Perceptions of Challenges When Pursuing a Career Path (upper bars) and Changes in Career Aspirations (lower bars)}
    \label{fig:RQ3}
\end{figure}

Figure~\ref{fig:RQ3} shows that most men (81.3\%) do not perceive gender-related barriers, compared to only 27.1\% of women. A small percentage of men (2.1\%) see such challenges, while 25.7\% of women do. Additionally, 47.1\% of women are unsure, while only 16.7\% of men share this uncertainty. This indicates that women are more likely to perceive or question gender-related challenges in their career paths. Regarding if respondents change their career aspirations, results show that a larger percentage of women (42.3\%) than men (36.2\%) have changed their career aspirations during university studies. Conversely, more men (44.7\%) than women (32.4\%) report no change. Additionally, 25.4\% of women and 19.1\% of men are unsure, indicating that women are slightly more likely to reconsider or remain uncertain about their career paths.

\subsection{Qualitative Findings}

Five women students from the final year of the Bachelor's programme in Software Engineering and Management agreed to participate in the interview. Interviews lasted an average of thirty minutes. The analysis resulted in three themes: Motivations for Choosing IT Programmes, Experiencing Gender Dynamics in IT Programmes at Gothenburg University and  Navigating Career Aspirations and Perceived Industry Challenges. We elaborate on them next.

\subsubsection{\textbf{Theme 1: Motivations for Choosing IT Programmes}}
This theme explores the motivations that drove students to choose IT-related programmes at Gothenburg University. These reasons included personal and structural factors influencing gender representation in these fields. A recurring factor across interviews was an early interest in technology and coding, often developed during high school or through prior exposure to programming. As one participant explained:

\dialoguegpt{quote}{
\pquote{5} "I was interested in computer science ever since high school and took courses in programming. I was interested in coding, so that was the main reason I was looking for a programme related to the computer science field".}

Another common motivation was that the programme combined technical and non-technical elements. The software engineering and management programme offers a balance of coding, management, and creative opportunities that particularly appealed to women students. Further, one important factor was the support and encouragement from family and, in some cases, even friends. This support was linked to the students' relatives' perception of IT as a promising and prestigious career field and being seen as a "modern" and "future-proof" profession. One participant remarked:

\dialoguegpt{quote}{
\pquote{1}"They (her parents) thought that software engineering is so good and that it is in high demand and pays really well. So my parents, both of them, pushed me towards trying it".}

These findings suggest that the design and accessibility of IT programmes, combined with early exposure to technology and societal support, are factors shaping women's choices in this field.

\subsubsection{\textbf{Theme 2: Experiencing Gender Dynamics in IT Programmes at Gothenburg University}}

This theme addresses women's experiences in GU's IT programmes. It shows a variety of inclusion, underrepresentation, and challenges of navigating gendered spaces. Most women reported positive experiences of equality and participation with balanced gender representation and active participation from female students. However, some others reported facing stereotypes, marginalisation, and feelings of isolation, pointing to systemic and cultural issues. They also noted gaps in technical content, such as limited mathematics courses, which might affect preparedness for advanced studies.
Among the students who felt included, the participant below expressed:

\dialoguegpt{quote}{
\pquote{5}"In my program, I think we had quite a lot of female students in our class. So that was a good thing to see. And many of the female students were participating in the lectures, and they were active students... everything was the same for both male and female students".}

This perception of inclusivity extended to classroom interactions and the overall learning experience.  

Contrary to this perception, other students expressed that the low number of women in IT programmes remains a significant challenge, influencing how they experience their academic environment. One participant reflected on this issue:

\dialoguegpt{quote}{
\pquote{4}"I generally feel sad about it, and a bit sorry for the women. When I see women in society, in a classroom full of, you know, guys, and very few girls".}

This quote shows that underrepresentation is not merely a statistical observation but also an emotional and psychological burden for this student. Additionally, one more student expressed that she has received comments about how she looks, which is linked to stereotypes.

\dialoguegpt{quote}{
\pquote{2}"I have received comments like, 'Oh, you don't look like someone who would do software engineering.' And that's just such a weird thing to say".}

This type of stereotyping challenges women's sense of belonging and reinforces traditional narratives about who is expected to pursue careers in IT. Beyond verbal comments, gender dynamics also manifest in group interactions. A participant shared her experience working with male classmates and not feeling listened. Although this participant actively sought ways to assert herself, the need to employ such strategies highlights the additional effort women often expend to be heard and respected in male-dominated spaces.

Faced with these challenges, women in IT programmes often develop resilience and adopt coping mechanisms to navigate gendered dynamics. One participant reflected on her approach to group work:

\dialoguegpt{quote}{
\pquote{3}"It's not like it's kind of affecting me drastically, because I'm also doing something to counteract that".}
Such resilience is important for their academic and personal success but also exposes a systemic issue: the responsibility to manage these dynamics often falls disproportionately on women themselves. While these strategies can mitigate immediate challenges, they do not address the root causes of inequality and exclusion.
In conclusion, these students' experiences show two narratives. Some aspects of IT programmes at Gothenburg University are inclusive and equitable. However, at the same time, there are still issues related to gender representation, stereotypes, and interaction dynamics.

\subsubsection{\textbf{Theme 3: Navigating Career Aspirations and Perceived Industry Challenges}}

This theme explores how university experiences and industry perceptions shape women's career aspirations. Participants expressed that exposure to diverse technical roles helped them discover areas of personal interest.

For example, P1 commented how specific coursework helped clarify their interests:
\vspace{10pt}

\dialoguegpt{quote}{
\pquote{1}"Then I had a lot of courses with UI/UX, and that is something that I realised that I enjoyed the most, so the programme helped me find my interests in regards to our opportunities as a software engineer".}

Conversely, another participant's experience in an internship exposed her to the underrepresentation of women in specific technical fields:

\dialoguegpt{quote}{
\pquote{5}"There were like only me as a female software engineer in that group. There were like 20 male developers and only one female. The group I was working with was working more with hardware and embedded software. I think that's not very interesting for female engineers in tech".}

This awareness of gender imbalance may not directly steer participants toward specific fields, but it does shape their perceptions of where they might feel more comfortable or supported. In this example, P5 connected this disparity to the technical nature of certain roles.

In conclusion, most participants chose their career paths based on their personal interests rather than overt gender biases. However, the underrepresentation of women in areas like hardware and embedded systems could influence their perceptions on this specific fields, even if indirectly.

\section{Discussion} \label{sec:discussion}
Below we discuss our findings and answer our RQs.

\subsection{RQ1: What factors influence gender representation among students in IT-related programmes?}

The survey showed that career prospects significantly influence decisions to pursue IT programmes, with similar proportions of women (32.1\%) and men (33.3\%) citing this as a motivating factor. However, interviews revealed that external validation often mediates women's motivations, such as family encouragement. As one participant noted, her parents emphasised the high demand and financial rewards of software engineering. This aligns with Ketchledge et al.~\cite{ketchledge2021factors} results, finding parents to be the second most influential factor in career choice and Bomanet al.'s~\cite{boman2024breaking} results, that encouragement from family and friends is a factor that also decreases social barriers to the sense of belonging in software engineering. Further, attention is drawn to the fact that while IT's status as a prestigious field can attract women, framing their involvement as an economic utility rather than interest may undermine long-term retention and satisfaction. Furthermore, the higher percentage of men citing "interest in technology/computing" (34.3\% vs. 25.5\% for women) can suggest that societal biases discourage women from viewing technology as an accessible or natural domain. This aligns with Cheryan et al.~\cite{cheryan2017some}, who argue that the masculine culture of STEM fields creates barriers to women's engagement.

Regarding how gender representation in programmes influences selection, 91.7\% of men and 80.3\% of women reported that gender distribution did not affect their decision; women were significantly more likely to prefer balanced or predominantly female cohorts. Only 14.1\% of women indicated they sought programmes with a balanced gender distribution, compared to only 8.3\% of men. While women's desire for balanced cohorts may seem pragmatic, it raises critical questions about the inclusivity of current IT education. In the interviews, women expressed wanting to have more girls around, and the balance was good for some others. Moreover, the lack of men expressing interest in programmes with more women might suggest that men may not consider diversity a factor in their educational choices. This reflects a broader societal trend where the burden of addressing gender imbalance falls disproportionately on women.

Concerning the amount of math and programming in the surveys, there is a clear difference between men and women: 58\% of men want more programming than 31\% of women. Additionally, 12.5\% of men said math was important in their programme choice, compared to 9.9\% of women, which aligned with Beyer's\cite{beyer2014women} results. However, in the interviews, all women commented on their desire to have more math in their programmes. 

Programmes that combine technical and non-technical elements— such as software engineering and management—were particularly attractive to women as they align with broader interests in creativity, collaboration, and problem-solving.

Relating to the influence of parents' education and occupations, male respondents were more likely to have fathers in technical or professional roles, potentially providing role models in IT. In contrast, women and non-binary respondents benefited from maternal influence, emphasising education and broader career possibilities. Regarding parents' education, the percentages have no significant differences (see Figure \ref{fig:parents} ). 

Men with fathers with Master's degrees or above were only 20,8\%, a slight difference from women's mothers (14,1\%). Similarly, both genders have similarly educated mothers, which might indicate that mothers' education has a more generalised influence on career choices rather than directly emphasising technical fields. These results contrast with Storen and Andersen's results \cite{storen2007women}, who found that fathers' higher education influences men's career choices in male-dominated fields and relate more with Brook's \cite{brooks2003young} findings on suggesting parental education has little impact on entry into technical fields like IT. 

\subsection{RQ2: How are women’s experiences in IT-related programmes
regarding gender?}

Women's experiences in IT-related programmes at Gothenburg University show a polarised picture. While some students report an inclusive and supportive environment, others experience marginalisation, stereotypes, and feelings of isolation. 

In the interviews, many women reported positive experiences and feeling included in classroom discussions and group work. For instance, one student expressed satisfaction with the balanced gender representation and active participation of female students. On the contrary, in the survey data, women's participation in classroom discussions and group projects was less than half; 39.4\% of women said they feel included "always" in discussions, compared to 55.3\% of men. This implies that gender balance in the classroom does not necessarily equate to a truly inclusive environment. Women may be present but not always given equal opportunities for participation or recognition, which could affect their sense of belonging and engagement in group work. Additionally, the prevalence of women reporting occasional or frequent biases— 21.4\% of women in the survey— indicates that these experiences of inclusion are not universal and that significant barriers remain. 

Furthermore, students also reported stereotypes, for example, the student who was told, \textit{"doesn't look like someone who would do software engineering"}, showing that stereotypes continue to be a prominent issue and could develop into bigger problems like academic performance \cite{nguyen2008does, schmader2008integrated}

Interestingly, the interview data also reveals a sense of resilience among women who navigate these challenges. One participant's reflection that \textit{"it's not like it's kind of affecting me drastically"} because she is actively counteracting the challenges points to a coping mechanism often necessary for women to succeed in such environments \cite{tokbaeva2023career}. However, this spots a critical systemic issue: the responsibility for managing gendered dynamics also falls on the women themselves rather than on the institution or male peers. The University of Gothenburg has developed several initiatives to attract, foster and retain women in IT fields. Those strategies also look to address the root causes of gender inequality and to lead to long-term change. These initiatives seek to create a pipeline for women into IT careers. However, the findings from this study indicate that more comprehensive and sustained efforts are needed. It is crucial to involve male peers in these efforts, ensuring that gender inclusivity is not solely the responsibility of women but is a shared goal across all students and faculty.

\subsection{RQ3: How does the university experience shape career aspirations and perceptions of gender-related challenges in IT fields?}

The university setting exposes women to new career opportunities and helps clarify their interests; it can also exhibit the challenges they face in male-dominated IT fields, influencing their career aspirations and shaping their perceptions of gender-based barriers.
Our survey results indicate that men are more likely to aspire to roles such as software development/engineering (38.2\%) compared to women (15.8\%), who show more interest in roles like UX/UI design (29.2\%) or IT project management (21.7\%). These findings aligned with Zhu's~\cite{zhu2019recoding} Beyer~\cite{beyer2014women} and Davie's~\cite{davies2005gender} results about ender and specific roles within the IT field.
In connection with the previously mentioned, one interviewee expressed that their exposure to certain technical fields, particularly hardware and embedded systems, reinforced gendered stereotypes about the roles suited to women. One interviewee shared an experience during an internship where she was the only female in a group of 20 male developers working with hardware: 

University programmes, particularly those with a broad curriculum, offer women opportunities to explore different IT-related career paths, allowing them to identify and pursue areas of personal interest. For instance, one interviewee noted that UI/UX design courses were particularly appealing and helped her discover her passion for this field: 

While the university environment exposes students to various career paths, it also influences their perceptions of gender-related barriers in IT. Survey data reveals that 25.7\% of women perceive gender-related challenges or barriers in pursuing their desired career path, compared to only 2.1\% of men. Furthermore, 47.1\% of women are unsure whether gender issues will impact their careers, reflecting the uncertainty and difficulties of navigating gendered dynamics in IT fields. The experiences shared in the interviews add to the picture. While some women felt empowered by their university experiences and confident in their career trajectories, others reported feeling isolated or marginalised in male-dominated spaces. 

It is imperative to look at results like those in the Figure \ref{fig:RQ3}  (upper bars). While 81.3\% of men perceived no gender-related barriers in pursuing their careers, only 27.1\% of women felt the same. As women continue to face barriers in men-dominated fields, the university's responsibility extends beyond simply fostering an inclusive curriculum. It must actively engage in dismantling the systemic gender biases that shape students' career choices and perceptions.

\subsection{Context Awareness}
This study was conducted in Sweden, a country widely recognised for its progressive stance on gender equality and ranked 1st in the EU on the Gender Equality Index~\cite{eige2024sweden}. The relatively high representation of women in some of the programmes we included might reflect the countries and university initiatives. Additionally, the topic usually attracts more minorities. However, despite the context, our findings indicate that gender-related challenges persist in IT education, a traditionally male-dominated field. 

The cultural context of Sweden offers a specific backdrop for exploring these dynamics, as the country's progressive policies might mask or reduce the visibility of subtler forms of gender bias. For instance, our study's prevalence of stereotypes and gendered barriers suggests that systemic challenges in IT education transcend national policies and cultural norms.

It is also crucial to recognise that the Swedish context may limit the generalisability of our findings to other cultural or educational environments. We discuss this in our threats to validity. In countries with less focus on gender equality, women may face even more pronounced barriers in IT education and careers. Conversely, in less progressive settings, the strategies proposed in this study might yield different outcomes or face more significant resistance. By situating our findings within Sweden's context, we aim to point at the importance of tailoring gender inclusivity efforts to specific cultural and institutional settings. While Sweden's policies and practices provide a valuable framework, addressing women's challenges in IT requires context-sensitive strategies that go beyond national-level indicators of equality.

\subsection{What Can We Do?}

Based on our findings and the actions we identify high-level actions that can be implemented in the classroom aiming to address the gender-related challenges and disparities. We acknowledge that context is a factor to take into account and that there are differences among universities. However, we think our proposal can be adapted to fit different settings.

\begin{itemize}
    \item \textbf{To foster interest-based career exploration} 
    
    Integrate opportunities for hands-on engagement with diverse technical fields, such as programming, hardware, and UX/UI design, to help students discover their passions.
    
    \item \textbf{To address gendered stereotypes} 
    
    Actively challenge stereotypes in course content and classroom interactions. Include discussions about gender biases in IT and highlight contributions from diverse professionals in traditionally male-dominated fields, such as hardware or software engineering.
    
    \item \textbf{To promote inclusive participation in classroom activities} 
    
    Use structured approaches in group discussions and projects to ensure all students, especially women, have equal opportunities to contribute.
    
    \item \textbf{To raise awareness and involve male peers in gender inclusion initiatives} 
    
    Develop programmes and integrate into specific classes topics to educate male students about the importance of gender inclusivity and allyship. Encourage shared responsibility in creating inclusive environments, reducing the disproportionate burden on women to navigate gendered dynamics.
\end{itemize}

\section{Threats to validity} 
In this section, we address the different validity threats that might limit our study's findings' reliability, generalisability, and accuracy.

\subsection{External Validity}

We acknowledge that our findings are strongly connected to the context since we only focus on students from one university. As a result, our findings cannot be assumed to be applicable or transferable to other study environments. We tried to mitigate this threat by including all students, given that most programmes in the IT faculty are international, a big sample of students come from different countries. Additionally, the findings in this study are focused on IT-related programmes which may limit the applicability to other fields of study where gender dynamics and stereotypes differ.
Finally, the small interview sample might not capture the full diversity of experiences, particularly among students from underrepresented cultural or socio-economic backgrounds. We correlate the findings with the survey, which has a bigger and more diverse sample to mitigate this threat. In addition, we also relate to general literature to frame our findings.

\subsection{Internal Validity}
Regarding the threats that affect internal validity, given the sensitivity of discussing gender issues, participants may have provided responses they believed were socially acceptable rather than fully honest. In addition, self-selection biases add to this threat; since participation was voluntary, students who participated in the survey and mainly in the interviews might be particularly motivated or have strong opinions about gender dynamics in IT. This could limit the representativeness of the findings. To mitigate this, we tried to reach many bachelor programme participants. Despite our efforts, we acknowledge that we received only five students willing to be interviewed, which may limit our study's depth and diversity of perspectives. To mitigate this, we ensured a thorough analysis of the data collected and triangulated the findings with survey responses to strengthen the validity of our conclusions.

Additionally, the researchers' cultural biases and assumptions introduced during the research process may similarly affect the conclusion validity of our study, much like confirmation bias. We tried to mitigate this by discussing our insights with the researcher's team. In the thematic analysis, two authors coded, and the third author acted as a reliable coder to avoid internal biases.

\subsection{Construct Validity}
In our data collection methods, we use several constructs, such as "inclusivity" and "gender-related biases", that could be open to different interpretations. To mitigate this, we ensured that the survey included clear response options that covered a range of experiences, and we used descriptive follow-up questions during interviews to help clarify participants' perceptions and align their answers with the intended meaning of the constructs.
\section{Conclussion} \label{sec:conclussion}
In this study, we investigate the factors that influence women and non-binary students to choose a university programme in the IT field at the University of Gothenburg using a combination of surveys and interviews. Additionally, we explore women's experiences at university in a male-dominated field and how this experience, if so, impacts their career aspirations. Moreover, we propose general actions to implement in the classroom to approach the disparities and challenges regarding gender disparities.

For future work, we aim to evaluate and analise specific actions taken in our university to improve gender challenges from the perspective of students. We also want to develop, implement and evaluate interventions based on our recommendations in the previous section. These interventions will mainly focus on retention by improving the university environment to support women and non-binary students.

\section{Acknowledgments}
We want to thank all the students who participated in this study, either through the survey or interview, as it would not have been possible to produce the research without their help.

\bibliographystyle{ACM-Reference-Format}
\balance
\bibliography{bibliography.bib}

\end{document}